\documentclass{paper}
\usepackage{epsfig, graphicx}
\usepackage[english]{babel}

\begin{document}
\title{On the Formation of a Macroscopically Flat Phospholipid Membrane
on a Hydrosol Substrate}
\author{\small Aleksey M. Tikhonov$^{\dagger}$\/\thanks{tikhonov@kapitza.ras.ru}, Viktor E. Asadchikov$^{\ddagger}$, Yuri O. Volkov$^{\ddagger}$}
\maketitle
\leftline{\it $^{\dagger}$Kapitza Institute for Physical Problems, Russian Academy of Sciences,}
\leftline{\it ul. Kosygina 2, Moscow, 119334, Russia}
\leftline{\it $^{\dagger}$Shubnikov Institute of Crystallography, Russian Academy of Sciences,}
\leftline{\it Leninskii pr. 59, Moscow, 119333 Russia}

\rightline{\today}

\abstract{The dependence of the structure of a phospholipid layer (DSPC
and SOPC) adsorbed on a hydrosol substrate on the concentration
of NaOH in a solution of 5-nm silica particles has been studied
by X-ray reflectrometry with the use of synchrotron radiation.
Profiles of the electron density (polarizability) have been
reconstructed from the experimental data within a
model-independent approach. According to these profiles, the
thickness of the lipid film can vary from a monolayer ($\sim 35$\,\AA
) to several bilayers ($\sim 450$\,\AA). At the volume
concentration of NaOH of $\sim$ 0.5 mol/L, the film on the hydrosol
surface is a macroscopically flat phospholipid membrane
(bilayer) with a thickness of $\sim 60$\,\AA{} and with areas of
$45 \pm 2$ and $49 \pm 3$\,$\,\AA^2$ per DSPC and SOPC molecule,
respectively.}

\vspace{0.25in}

\large

A phospholipid bilayer is the simplest model of a
cell membrane \cite{1,2,3}. The method of obtaining macroscopically
flat multilayers of lipid membranes on a
strongly polarized substrate of an aqueous solution of
amorphous silica nanoparticles was proposed in our
work [4]. Using a model-independent approach to
reconstruct electron density profiles [5, 6], we
revealed from the X-ray reflectometry data that the
thickness of a phospholipid multilayer is determined
by the following parameters of the hydrosol substrate,
which specify the width of the surface electric double
layer: the concentration of Na$^+$, pH level, and size of
nanoparticles. In particular, a macroscopically flat
phospholipid membrane spontaneously appears on
the surface of the hydrosol of 5-nm particles under
certain conditions.

Films of 1,2-distearoyl-sn-glycero-3-phosphocholine
(DSPC) and 1-stearoyl-2-oleoyl-sn-glycero-3-
phosphocholine (SOPC, see Fig. 1) were prepared
and studied with the methodology described in [4]. A
10- to 20-$\mu$\,L drop of a solution of phospholipid in
chloroform was deposited from a syringe on the surface
of the liquid substrate placed on a fluoroplastic
plate with a diameter of $\sim 100$\,mm; the amount of substance
in the drop was enough for the formation of
more than ten lipid monolayers after its spreading on
the surface. In this case, the adsorbed film is in equilibrium
with three-dimensional aggregates in which an
excess of the surfactant is accumulated. A change in
the surface tension ã of the air–hydrosol interface
from $\sim 74$ to $\sim 50 – 30$\,mN/m was detected by the Wilhelmy
method using an NIMA PS-2 surface pressure
sensor. Then, the equilibrium of the sample was
reached inside a hermetic single-stage thermostat at
T = 298\,K in about 12\,h.

Powders of synthetic of synthetic DSPC and
SOPC and their solutions in chloroform was purchased
from Avanti Polar Lipids and chloroform
($\sim 99.8\%$) was purchased from Sigma-Aldrich. The
hydrophobic part of molecules of these lipids ($L_1 \approx
2$\,nm) consists of two hydrocarbon chains including
18 carbon atoms. The hydrophilic part of the molecule
($L_2\approx 1.5$\,nm) is formed by glycerin and phosphocholine.
The only difference between the structures of
DSPC and SOPC is the presence of a double carbon
bond between the ninth and tenth atoms in one of the
hydrocarbon chains in the latter molecule. The DSPC
(or C$_{44}$H$_{88}$NO$_8$P) and SOPC (or C$_{44}$H$_{86}$NO$_8$P) molecules
contain $\Gamma = 438$ and 436 electrons, respectively.

\begin{figure}
\hspace{0.5in}
\epsfig{file=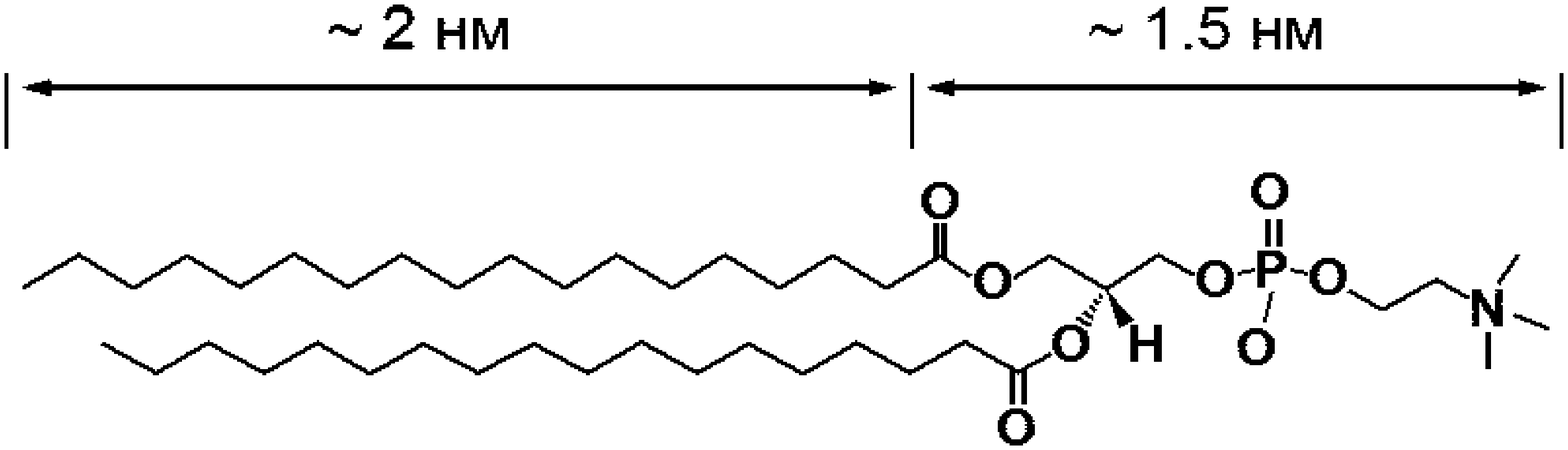, width=0.75\textwidth}

Figure 1. Molecular structure of the DSPC phospholipid.
\end{figure}

Concentrated monodisperse Ludox FM sol stabilized
by sodium hydroxide was obtained from the
Grace Davison Co. This aqueous solution with a density of 1.1 g/cm$^3$ 
includes amorphous silica particles
with the diameter $D \approx 5$\,nm (16 wt\% SiO$_2$, 0.3 wt \%Na, and pH$ \approx 10$).

The  hydrosol was enriched by NaOH in a vessel by its
mixing (shaking and subsequent deposition in a Branson
2510 ultrasonic cleaner) with a solution
($\sim 5$\,mol/L) of alkali metal oxide (99.95\% of the metal,
Sigma-Aldrich) in deionized water (Barnstead UV). It
is very important that the pH level of sol not exceed the
critical value pH$_c$ < 12, at which coagulation of
nanoparticles occurs \cite{7}.

The surface-normal structure of lipid films was studied by the
reflectometry method with the use of synchrotron
radiation at the X19C station of the NSLS synchrotron,
which was equipped with a universal spectrometer
for study of the surface of a liquid \cite{8}. A bending
magnet with a critical energy of $\sim 6$\,keV was used as a
source of radiation for the X19C station. A focused
monochromatic X-ray beam with an intensity of
$\approx 10^{11}$ photon/s and an energy of photons $E = 15$\,keV
($\lambda = (0.825 ± 0.002)$\,\AA) was used in the experiments.
The beam whose cross section at the output of the
magnet had a height of 5 mm and a width of 40 mm
was focused by a toroidal mirror with a focal length of
$\sim 10$\,m. Then, the beam with the diameter of the cross
section less than 1 mm was deflected by means of a
single-crystal (Si (111)) monochromator to the surface
of the sample oriented by the gravitational force. Thus,
the range of the glancing angle $\alpha$ from $0^\circ$ to $\sim 8^\circ$ could
be covered when measuring the reflection coefficient.
The monochromator of the spectrometer was based on
a three-circle goniometer (Huber), had water cooling,
and was placed in a hermetic chamber filled with gaseous
helium at a low excess pressure ($\sim 10$\,Torr). The
geometric parameters of the beam incident on the surface
of the sample, as well as the spatial resolution of
the detector, were controlled in the experiments by
means of slits. In this work, the reflection coefficient
was measured by a detector with the angular resolutions
$\Delta\beta\approx 0.02^\circ$ and $0.8^\circ$ in the vertical and horizontal
planes, respectively.

Let {\bf k}$_{\rm in}$ and {\bf k}$_{\rm sc}$ be the wave vectors of the incident
beam and the beam scattered in the direction to the
observation point, respectively. The scattering vector
{\bf q = k$_{\rm in}$ {\rm -} k$_{\rm sc}$} at mirror reflection ($\alpha=\beta$) has only one
component $q_z=(4\pi/\lambda)\sin(\alpha)$, where $\alpha$ and $\beta$ are the
glancing and scattering angles, respectively, in the
plane normal to the surface (see inset in Fig. 2).

When measuring the reflection curve, the effect of
lateral inhomogeneities of the surface and near-surface
layers is averaged because the characteristic area
of the region illuminated by a probe beam on the sample
is $\sim 100$ mm$^2$. As a result, the reconstructed structures
can be considered within the notion of an ideal
layered inhomogeneous medium.

\begin{figure}
\hspace{0.5in}
\epsfig{file=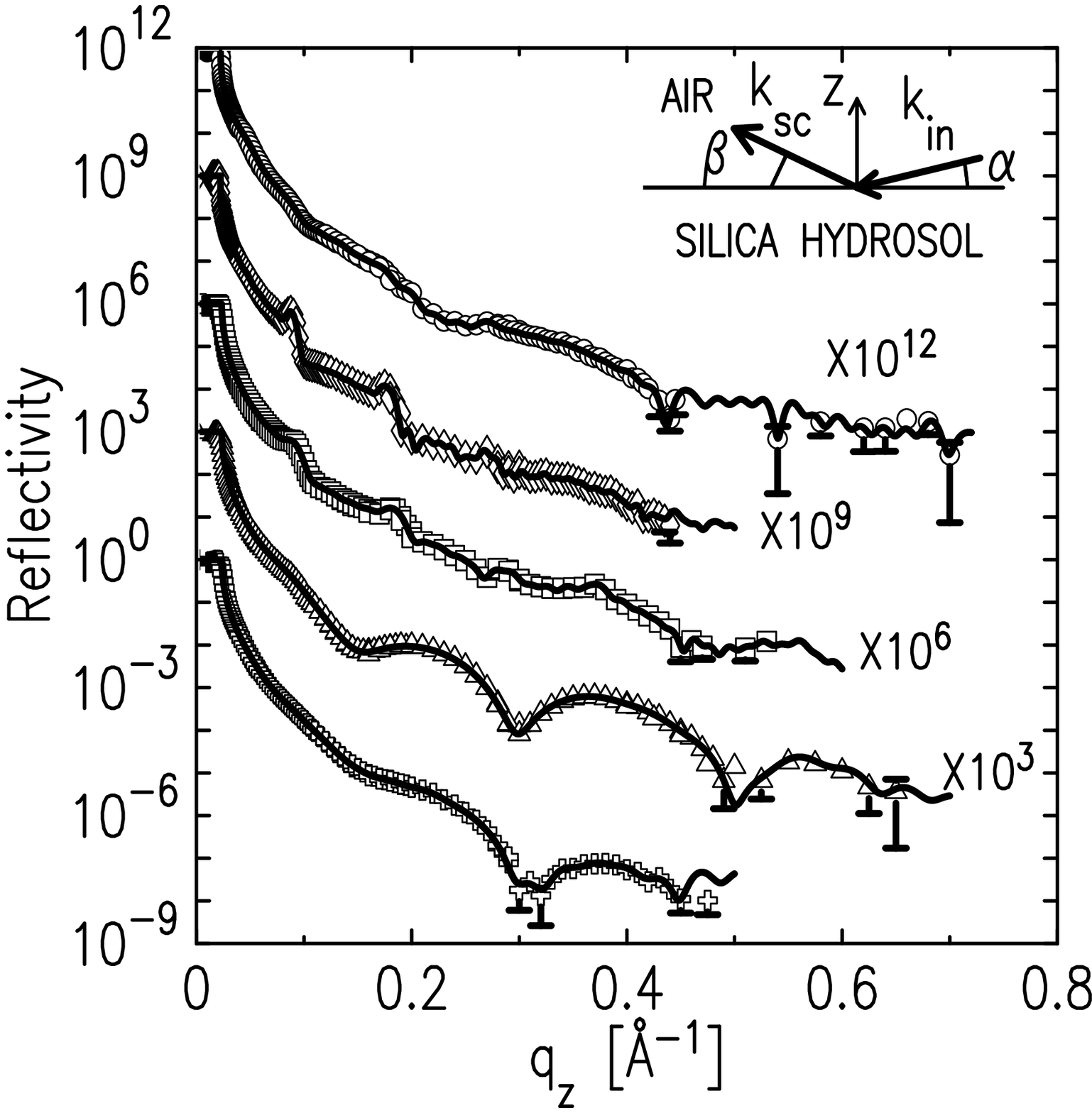, width=0.75\textwidth}

Figure 2. Reflection coefficient $R$ of the air–hydrosol interface
with the adsorbed lipid film for (circles) the DSPC
monolayer (pH\,$\approx10$); (diamonds) the DSPC multilayer
pH\,$\approx10$); (squares) the DSPC multilayer pH\,$\approx11$); (triangles
and crosses) the DSPC and SOPC bilayers, respectively
pH\,$\approx11.5$); and (solid lines) calculations. The inset
shows the wave vectors {\bf k}$_{\rm in}$ and {\bf k}$_{\rm sc}$ of the incident beam and
the beam scattered in the direction to the observation
point, respectively.
\end{figure}

The detailed reconstruction of the distributions of
the polarizability of the medium $\delta(z)$ over the depth $z$
was performed within a model-independent approach
based on the extrapolation of the asymptotic angular
dependence of the reflection coefficient to the region
of large $\alpha$ values \cite{5,6}. It is assumed that the reconstructed
structure includes peculiar "points of discontinuity"
at which either $\delta(z)$ or its nth derivative $\delta^{(n)}(z)$
changes stepwise:
\begin{equation}
  \label{eq:discontin}
  {\rm D}^{(n)}\left(z_j\right) \equiv
  \frac{d^n\delta}{dz^n}\left(z_j + 0\right)
  -\frac{d^n\delta}{dz^n}\left(z_j - 0\right),
\end{equation}
where $z_j$ is the coordinate of the $j$th point of discontinuity.
In turn, the asymptotic behavior of the amplitude
reflection coefficient in the first Born approximation
has the form
\begin{equation}
  \label{eq:fba}
  r\left(q_z\to\infty\right) \simeq
  -\left(\frac{2\pi}{\lambda}\right)^2\left(\frac{i}{q_z}\right)^{n+2}
  \sum\limits_{j=1}^m {\rm D}^{(n)}\left(z_j\right)e^{iq_zz_j}.
\end{equation}

According to \cite{5}, a finite number of different amplitude
reflection coefficients describing the experimentally
measured square of their absolute value $R$ in a
certain interval of $q_z$ correspond to a given combination
of $m$ points of discontinuity of ${\rm D}^{(n)}\left(z\right)$. In particular,
if the distances between all points of discontinuity
are different, there are only two solutions $\delta(z)$ satisfying
the required asymptotic behavior of the reflection
curve.

The procedure of the model-independent reconstruction
of the polarizability profile includes two
stages. First, the order and positions of the points of
discontinuity for the structure under study are determined
by analyzing the dependence $R\cdot q_z^{2n+4}$ (where $n = 0,1,2,\ldots$ is the desired order of singular points). Then,
the distribution $\delta(z_1,\ldots,z_M)$ divided into a large number
$M\sim 100$ of thin layers is numerically optimized. In
this case, the calculated reflection curve $R_c$ is fitted to
the experimentally measured curve $R$ with the use of
the standard Levenberg–Marquardt algorithm \cite{9}.

All experimental curves in this work decrease as
$\propto 1/q_z^6$. Consequently, to describe the structures, it is
sufficient to use only the singular points of the first
order. In this case, the residual target function ensuring
the required asymptotic behavior of the angular
dependence of the reflection coefficient has the form
\begin{equation}
\begin{array}{lllllll}
  \label{eq:minfun}
    MF\left(\delta_1,\ldots,\delta_M\right) = \\ \\
=\frac{1}{N}\sum\limits_{j=1}^N\Bigl[\log R\left(q_{j}\right) - \log R_{c}\left(q_{j}\right)\Bigr]^2+ \\ \\
+ Q_1\sum\limits_{j\neq j_1,\ldots ,j_m}^{M-1}\left(\delta_{j-1}+\delta_{j+1}-2\delta_j\right)^2+ \\ \\
+ Q_2\sum\limits_{j=j_1\ldots j_m}^{m}\left(\delta_{j+1}-\delta_j\right),
\end{array}
\end{equation}
\noindent
where $N$ is the number of experimental points; $j_1\ldots j_m$
are the positions of the points of discontinuity; and
$Q_{1,2}\approx 10^9$ are the parameters controlling the accuracy
of fitting. The second sum in Eq. (3) ensures the continuity
of the profile $\delta(z)$ in the intervals between the
points of discontinuity $z_{1}\ldots z_{n}$, and the third sum
ensures the first order of points of discontinuity.

The reconstructed profile $\delta(z)$ corresponds to the
electron density distribution $\rho(z)$ \cite{10}:
\begin{equation}
  \label{eq:polariz}
  \rho=\frac{2\pi}{r_0\lambda^2}\delta
\end{equation}
where $r_0=2.814\times 10^{-5}$\,\AA{} is the classical radius of the
electron. Then, the dependence $\rho(z)$ can be used, e.g.,
to estimate the area $A$ per molecule in a monolayer
with the thickness $d=z_2-z_1$:
\begin{equation}
  \label{eq:surf}
  A =\frac{\Gamma}{\int\limits_{z1}^{z2}\rho(z)dz},
\end{equation}

Figure 2 shows the experimental dependences of
the reflection coefficient $R(q_z)$ for the interfaces
between air and the hydrosol of 5-nm particles with
the adsorbed lipid film. Circles correspond to the surface
of the sol with ${\rm pH} \approx 10$ (Ludox FM) where the
amount of deposited DSPC is insufficient for the formation
of a uniform monolayer on the entire surface of
the substrate. The period of oscillation of R for this
system is $\Delta q_z \approx 0.25$\,\AA $^{-1}$, which implies the presence of
an adsorbed monolayer ( $\sim 2\pi/\Delta q_z\approx 30$\,\AA) in the illumination
region. Diamonds show the dependence
$R(q_z)$ for the same surface of the sol with ${\rm pH} \approx 10$ but
with a homogeneous DSPC multilayer. Squares correspond
the surface of the NaOH-enriched
($\sim 0.3$\,mol/L) sol with ${\rm pH}\approx 11$ with the same phospholipid
multilayer. The last two dependences are
similar to the previously reported data for hydrosol
substrate with silica particles with a diameter of
$\sim 22$\,íì and ${\rm pH}\approx 9$ \cite{4}. The data presented by triangles
(DSPC layer) and crosses (SOPC layer) were
obtained for NaOH-enriched ($\sim 0.5$\,mol/L) substrates
with ${\rm pH} \approx 11.5$. The period of oscillations $R$ for these
data is $\Delta q_z$ $\approx 0.15$\,\AA$^{-1}$; i.e., the thickness of the
adsorbed layer $\sim 2\pi/\Delta q_z\approx 50$\,\AA{} (bilayer).

According to the $\rho(z)$ profile reconstructed for the
DSPC monolayer (Fig. 3a, where $\rho_w=0.333$  {\it e$^-$/}{\AA}$^3$ is
the electron density in water under normal conditions),
its thickness is $36\pm 2$~\AA{} $\approx L_1+L_2$; i.e., all
molecules are predominantly oriented along the normal
to the surface. Although the calculated area per
molecule $A = 44\pm 2$~\AA$^2$ is in good agreement with
the value for the crystal monolayer, the electron density
in the region of the hydrophilic group is lower than
that in the model distribution (dashed line in Fig. 3à)
for the DSPC monolayer from \cite{4}. This likely indicates
the incomplete filling of the lipid layer.

\begin{figure}
\hspace{0.5in}
\epsfig{file=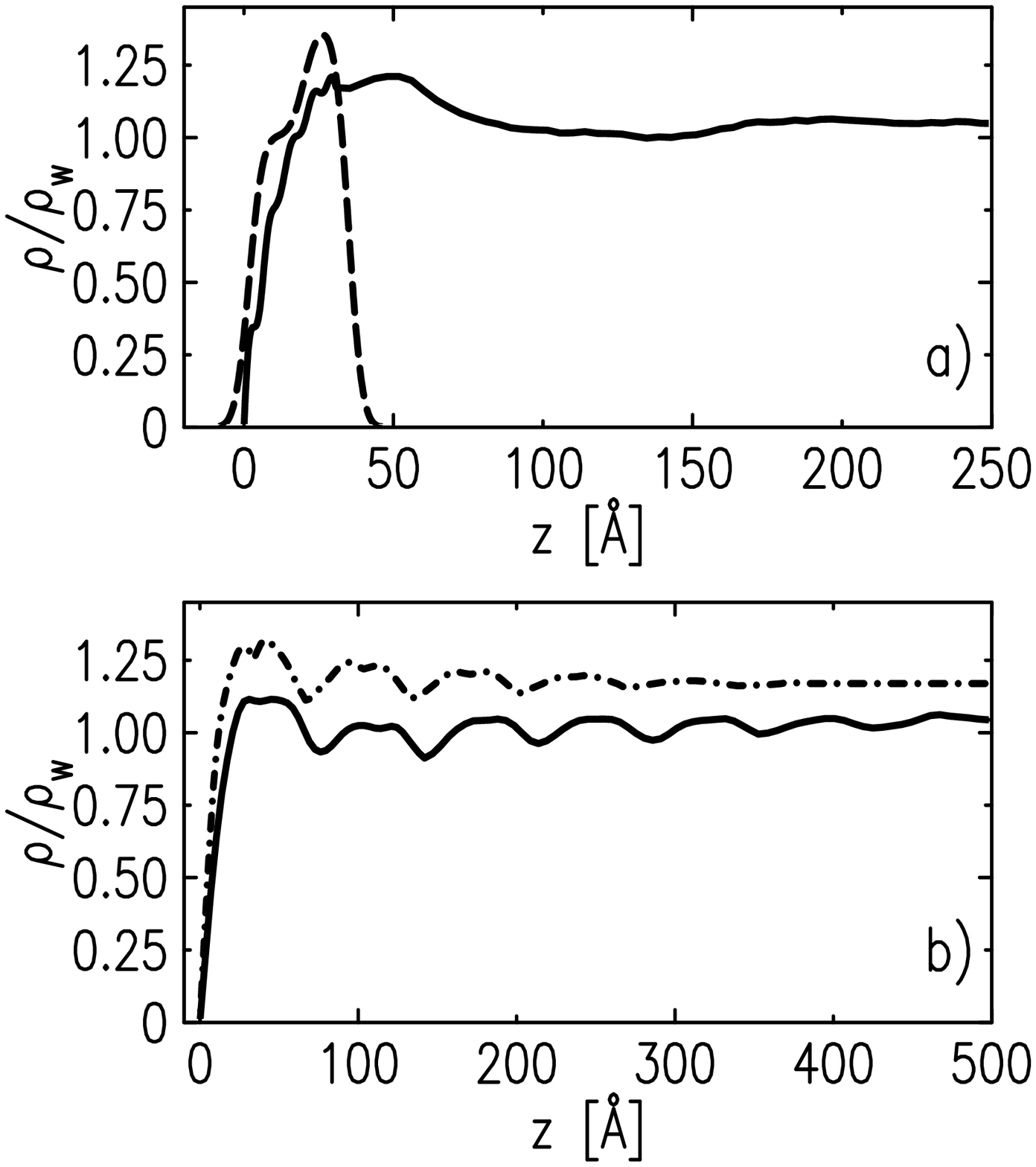, width=0.75\textwidth}

Figure 3. Reconstructed distribution profiles $\rho(z)$ normalized
to the electron density in water under normal conditions,
$\rho_w=0.333$  {\it e$^-$/}{\AA}$^3$. (a) The solid line is for the DSPC
monolayer on the surface of the hydrosol of $\sim 5$-nm particles
and the dashed line is the model electron density distribution
for the DSPC monolayer \cite{4}. (b) The solid line is
for the DSPC multilayer on the hydrosol substrate with
${\rm pH}\approx10$ and dash-dotted line is for the same lipid multilayer
on the substrate with ${\rm pH}\approx 11$.
\end{figure}

The layer with an increased density $\rho \approx 1.2 \rho_w$ with
the thickness $\approx 50$~\AA{} directly follows the monolayer
and the next layer has a thickness of $\sim 80$\AA{} and an
electron density of $\approx \rho_w$. We attribute the formation of
the last two layers to the condensation of silica
nanoparticles at the boundary of the monolayer
formed by hydrophilic groups \cite{11}.

The profile of the DSPC multilayer on the surface
of the hydrosol with ${\rm pH} \approx 10$ (solid line in Fig. 3b) has
a six-layer structure (total thickness $\sim 450$~\AA) with the
period between the points of discontinuity of $(72\pm 2)$~\AA,
which corresponds to the doubled length of the DSPC
molecule. Assuming that each of the observed layers is
equivalent to a molecular bilayer, we calculated the
value $A = (39\pm1)$~\AA$^2$. At the same time, according to
the experimental data on grazing diffraction, $A=(41.6\pm0.7)$~\AA$^2$ 
\cite{4}. Thus, according to Eq. (5), the
excess number of electrons per lipid molecule is $32\pm 9$,
 which corresponds to about three H$_2$O molecules
and/or Na$^+$ ions. The thickness of the DSPC multilayer
on the substrate with ${\rm pH} \approx11$  is noticeably
smaller ($\sim 300$~\AA{}, the dash-dotted line in Fig. 3b). This
multilayer has a four-layer structure with the period
between the points of discontinuity of $(68.1 \pm 0.9)$\,\AA{}
and with the calculated value $A = (34 \pm 2)$~\AA$^2$. Thus,
the "excess" number of electrons per lipid molecule is
$86\pm 2$, which corresponds to nine H$_2$O molecules and/or
Na$^+$ ions.

Figure 4à shows the electron density profile for a
thin film of the DSPC lipid on hydrosol heavily
enriched in NaOH ($\sim 0.5$\,mol/L, ${\rm pH} \approx 11.5$). One of
the points of discontinuity in it is located at a depth of
30.6~\AA. Assuming that the position of this singular
point corresponds to the interface between the outer
and inner monolayers, we obtain $A =45 \pm 2$~\AA$^2$.

The denser layer adjacent to this interface has a
thickness of $\sim 20$~\AA. On one hand, its thickness is
slightly smaller than that of the DSPC monolayer and
the total density is close to the total density of the latter
monolayer. On the other hand, the thickness of this
layer is less than half of the characteristic diameter of
particles in the volume of the initial sol ($\approx 5$~nm). If this
layer is formed by colloid particles, it is necessary to
assume that their radius decreases significantly with
an increase in the concentration of sodium in a solution.
However, at a high concentration of Na$^+$, an
inverse process—coagulation of particles, which is
manifested, e.g., in the turbidity of the solution—was
experimentally observed \cite{7}. This allows the interpretation
of this distribution as a profile of the lipid
bilayer on the surface of the hydrosol.

For the SOPC lipid film on the substrate with
${\rm pH} \approx 11.5$, the thickness of the dense region is no more
than 60~\AA{} (Fig. 4b). In this case, $\rho(z)$ decreases
smoothly with the depth and, correspondingly, no
interface between the film and substrate is observed.
No pronounced stratification of silica nanoparticles in
the surface region is observed either. The total density
of the entire observed structure at depths up to 57~\AA{} is
more than twice as large as the theoretical value for the
monolayer of this lipid. Under the assumption that the
amount of SOPC in the film corresponds to two
(three) monolayers, we obtain $A = (49 \pm 3)$\,\AA$^2$ ($= (65 \pm 3)$~\AA$^2$).
Only the latter value is in agreement
with the estimate of the quantity A for bilayer walls of
SOPC vesicles in aqueous suspensions \cite{12}.

\begin{figure}
\hspace{0.5in}
\epsfig{file=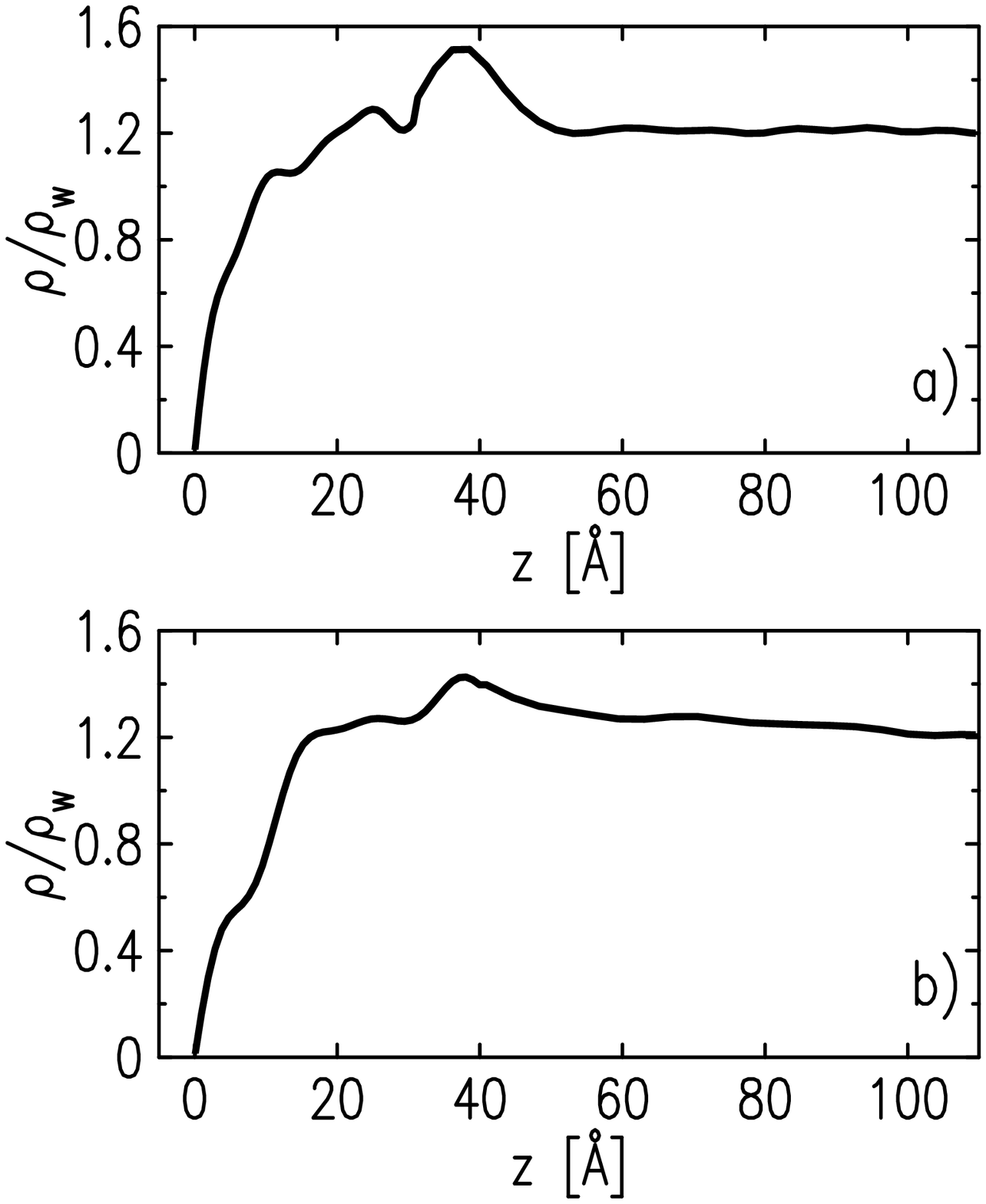, width=0.75\textwidth}

Figure 4. Reconstructed distribution profiles $\rho(z)$ normalized
to the electron density in water under normal conditions,
$\rho_w=0.333$  {\it e$^-$/}{\AA}$^3$, for the (a) DSPC and (b) SOPC membranes
on the substrate with ${\rm pH}\approx 11.5$.
\end{figure}

Electron density profiles for the DSPC and SOPC
films in Fig. 4 have a very similar structure. At the
same time, DSPC and SOPC phospholipids have different
temperatures $T_c$ of a phase transition associated
with the melting of hydrocarbon chains (chain-melting
transition) \cite{2}. For the former phospholipid, $T_c \approx 55$\,$^{\circ}$C, which explains the formation of the macroscopically
flat crystalline membrane ($A = (45\pm 2)$\,\AA$^2$). For
SOPC, $T_c \approx 6$\,$^{\circ}$C and the parameter $A$ is $\sim 10\%$ larger
than that for DSPC. Thus, the SOPC membrane at
room temperature is possibly in a liquid aggregate
state.

For the last two systems, the interval $z$ in which the
surface electron density differs from the bulk density,
$\rho_b \approx 1.2\rho_w$, is $50-60$\,\AA wider than the thickness of the
bilayer ($\sim 60$\,\AA). The appearance of this transient
region is possibly due to the condensation of silica
nanoparticles at the edge of the lipid membrane \cite{11}.

It was previously shown that a very wide transient
layer (electric double layer) appearing because of the
difference between the potentials of the forces of the
electric image for Na$^+$ cations and negatively charged
silica nanoparticles (macroions) exists at the (air–silica
sol) interface \cite{13}. Its width is determined by the
Debye screening length $\Lambda_D$ in the bulk of the solution
\cite{1}. The addition of sodium hydroxide to the composition
of the sol results in the shift of chemical equilibrium
in it, which is accompanied by an increase in the
pH value of the solution or the volume concentration
$c^-$ of OH$^-$ ions. Since $\Lambda_D\propto1/\sqrt{c^-}$, an increase in pH
leads to the narrowing of the double surface layer \cite{14}.

\vspace{7mm}
\small
{\bf Table}. Parameters of structures on the surface of the NaOH-stabilized sol of 5-nm SiO2 particles: the thickness L and the surface area A per lipid molecule.

\vspace{2mm}
\hspace{5mm}
\begin{tabular}{|c|c|c|c|c|}
\hline
&&&&\\
Structure & Phospholipid  & pH & $L$ ({\AA}) & $A$ ({\AA}$^2$) \\
&&&& \\
\hline
&&&& \\
Monolayer & DSPC  & 10 &$ 36 \pm 2 $  &  $44  \pm 2 $\\
&&&& \\
\hline
&&&& \\
Bilayer in a multilayer & DSPC  & 10 & $ 72 \pm 2 $   &  $41.6 \pm 0.7$\\
&&&& \\
\hline
&&&& \\
Bilayer in a multilayer & DSPC  & 11 & $ 68 \pm 1 $   &  $41.6 \pm 0.7$\\
&&&& \\
\hline
&&&& \\
Bilayer & DSPC & 11.5 &$ 60 \pm 2 $  &  $45 \pm 2 $\\
&&&& \\
\hline
&&&& \\
Bilayer & SOPC & 11.5 &$ 60 \pm 2 $  &  $49  \pm 3 $\\
&&&& \\
\hline

\end{tabular}
\large
\vspace{7mm}

The set of our data indicates that the total thickness
of the adsorbed DSPC multilayer is $\sim \Lambda_D$ (see table).
The enrichment of the hydrosol substrate with NaOH
results in a decrease by several times in the maximum
thickness of the adsorbed lipid layer according to a
decrease in $\Lambda_D$. The thickest ($\sim 450$\,\AA) multilayer of six
DSPC bilayers is formed on the surface of the hydrosol
at ${\rm pH} \approx 10$ ($\Lambda_D\sim 300$\AA), the multilayer with a
thickness of $\sim 300$\,\AA{}  at ${\rm pH} \approx 11$ ($\Lambda_D\sim 100$\AA) consists
of four bilayers, and one bilayer appears at ${\rm pH} \approx 11.5$ ($\Lambda_D\sim 50$\AA).
In this case, the oriented packing of molecules
inside each bilayer corresponds to a two-dimensional
phospholipid crystal.

To summarize, electron density profiles reconstructed
within the model-independent approach
demonstrate that the thickness of the DSPC film
adsorbed on the surface of the hydrosol coincides in
order of magnitude with the Debye screening length in
the substrate. At the volume concentration of NaOH
$\sim 0.5$\,mol/L and pH\,$\approx 11.5$, a macroscopically flat
phospholipid membrane with a thickness of $\sim 60$\,\AA{} and
with the $A$ value characteristic of a two-dimensional
crystal (hexagonal phase $P_{\beta '}$ \cite{2,4}) is formed on the
surface of the hydrosol. The $A$ value for SOPC is at
least $\sim 10\%$ larger than that for the DSPC bilayer. In
this case, the formation of a liquid membrane with a
thickness of $\sim 60$\,\AA{} (phase $L_\alpha$ \cite{2}) is possible.

Use of the National Synchrotron Light Source, Brookhaven National Laboratory, was supported by the U.S. Department of Energy, Office of Science, Office of Basic Energy Sciences, under Contract No. DE-AC02-98CH10886.
The operation of the X19C station was supported by the ChemMatCARS Foundations
of the University of Chicago, the University of Illinois at Chicago, and Stony Brook University.
The author also thanks Grace Davison for providing Ludox solutions of colloidal silica.
This work was supported in part by the Russian
Foundation for Basic Research (project no. 15-32-20935).


\small
\begin{thebibliography}{29}
\bibitem{1}
A.\,W.\,Adamson, {\it Physical Chemistry of Surfaces}, 3rd ed. John Wiley \& Sons: New York, 1976.
\bibitem{2}
D.\,M.\,Small, {\it The Physical Chemistry of Lipids}, Plenum Press, New York, 1986.
\bibitem{3}
D.\,A.\,Los$^\prime$,Fatty Acids Desaturases (Nauchn. Mir, Moscow, 2014) [in Russian].
\bibitem{4}
A.\,M.\,Tikhonov, JETP Lett. 92, 356 (2010); arXiv:1010.1680v1 [cond-mat.soft].
\bibitem{5}
I.\,V.\,Kozhevnikov,  Nuclear Instruments and Methods in Physics Research A 508, 519 (2003).
\bibitem{6}
I.\,V.\,Kozhevnikov and L.\,Peverini and E.\,Ziegler, Phys. Rev. B 85, 125439 (2012).
\bibitem{7}
J.\,Depasse, A.\,Watillon, J. Colloid and Interface Sci. 33, 430 (1970).
\bibitem{8}
M.\,L.\,Schlossman, D.\,Synal, Y.\,Guan, M.\,Meron, G.\,Shea-McCarthy, Z.\,Huang, A.\,Acero, S.\,M.\,Williams, S.\,A.\,Rice, P.\,J.\,Viccaro, Rev. Sci. Instrum. 68, 4372 (1997).
\bibitem{9}
J.\,Nocedal, S.\,Wright, {\it Numerical Optimization}, 2nd ed., Springer, 2006.
\bibitem{10}
B.\,L.\,Henke, E.\,M.\,Gullikson, J.\,C.\,Davis, Atomic Data and Nuclear Data Tables, 54, 181 (1993).
\bibitem{11}
V.\,E.\,Asadchikov, V.\,V.\,Volkov, Yu.\,O.\,Volkov, K.\,A.\,Dembo,
I.\,V.\,Kozhevnikov, B.\,S.\,Roshchin, D.\,A.\,Frolov,
and A.\,M.\,Tikhonov, JETP Lett. 94, 585 (2011); arXiv:1111.0955v1 [cond-mat.soft].
\bibitem{12}
N.\,Ku\v{c}erka, M.-P.\,Mieh and J.\,Katsaras, Biochimica and Biophysica Acta 1808, 2761 (2011).
\bibitem{13}
A.\,M.\,Tikhonov, J. Phys. Chem. C 111, 930 (2007).
\bibitem{14}
A.\,M.\,Tikhonov, J. Chem. Phys 124, 164704 (2006).

\end{thebibliography}
\end{document}